\documentclass[pra,twocolumn,a4paper,showpacs,amsmath]{revtex4}
\usepackage{graphicx}
\usepackage{bm} % bold math

\newcommand{\nn}{\nonumber}

\begin{document}
\title{Precessional motion of a vortex in a finite-temperature Bose-Einstein condensate}

\date{\today}

\author{Tomoya Isoshima}
\email{tomoya@focus.hut.fi}
\author{Jukka Huhtam\"{a}ki}
\author{Martti M. Salomaa}

\affiliation{
Materials Physics Laboratory,
Helsinki University of Technology, \\
P.~O.~Box 2200 (Technical Physics), FIN-02015 HUT, Finland
}

\begin{abstract}
We study the precessing motion of a vortex in a Bose-Einstein condensate
of atomic gases.
In addition to the former zero-temperature studies, 
finite temperature systems are treated
within the Popov and semiclassical approximations.
Precessing vortices are discussed utilizing the rotating frame of reference.
The relationship between the sign of the lowest excitation energy and
the direction of precession is discussed in detail.
\end{abstract}

% 03.75.Lm Tunneling, Josephson effect,
%     Bose-Einstein condensates in periodic potentials, 
%     solitons, vortices and topological excitations
% 03.75.Kk 
%     Dynamic properties of condensates;
%     collective and hydrodynamic excitations, superfluid flow
% 05.30.Jp Boson systems
% 67.40.Vs Vortices and turbulence

\maketitle

\section{Introduction}

The Gross-Pitaevskii (GP) approximation is often used to
analyze the Bose-Einstein condensates (BEC) of atomic gases~\cite{pethick}.
The GP approximation only treats the condensate fraction,
leaving out the normal component.
Therefore, the GP equation treats systems at zero temperature.
The Bogoliubov equations~\cite{pethick} describe the excitations
supported by the condensate in the zero-temperature limit.
The Bogoliubov excitation spectra agree with experimental results,
such as 
the collective oscillation modes (including the Tkachenko waves)
\cite{baksmaty0307368,mizushima} of a vortex lattice and
those on Bragg spectroscopy \cite{tozzo} of elongated condensates.
Since the physical systems exist at finite temperatures,
attempts to incorporate effects of temperature into the various approximations
have been made~\cite{projectedGP,ZGN}.
Among these, the Hartree-Fock-Bogoliubov-Popov (Popov)
approximation (Sec.~V.~F in Ref.~\cite{review:dalfovo}) is one of the most
common finite-temperature approximations.

The Popov approximation treats the density of the normal gas component as a mean field.
The condensate is described with the GP equation, extended to
include the mean field potential.
The excitation spectra and the wavefunctions of the normal gas
are given by eigenequations which are similar to the Bogoliubov equations.
The excitation spectra and the wavefunctions constitute the density of the normal gas. 

Within the Bogoliubov ($T=0$) approximation,
the occurrence and the disappearance of a vortex are indicated by the existence of
an excitation with negative excitation energy (in other words, the anomalous excitation).
There also exist relations between the direction of the precessional motion
of the vortex and
the sign of the excitation energies within zero-temperature theories.
This paper aims to answer the following question:
Does this relation stay valid in the Popov ($T>0$) approximation?

Besides the problem of the direction of precession,
the precession frequency within the Popov approximation has not been investigated in detail.
The precession frequency at zero temperature has been discussed 
within the Thomas-Fermi (TF) approximation using several methods
(see Sec. 5.1 of Ref. \cite{fetter:review} and Ref. \cite{lundh:ao}).
There also exists an analysis within the classical-field approximation~\cite{schmidt}
in finite-temperature systems.

The condensate is assumed to be trapped with a rotationally symmetric trap.
Therefore, a precessing vortex is an off-centered vortex.
Within zero-temperature theories,
a static system with an off-centered vortex~\cite{jukka} in the rotating frame
is equivalent to a system in the presence of a precessing vortex.
If the displacement $\Delta r$  of the vortex from the trap center is small,
the excitation energy of the anomalous mode~\cite{linn}, or the core-localized mode
depends linearly on the rotation frequency.
The variational Lagrangian analysis used in Ref.~\cite{linn}  and
Bogoliubov theory yield the same dependence for a centered vortex.

In a previous study~\cite{sami},
we discussed a precessing vortex state as a deviation from the axisymmetric configuration.
In this work we aim to treat the precessing vortex
in a 2D geometry directly within
the Popov approximation
in order that the relation between the precession and the excitations are
directly taken into account from the outset.

\section{approximations}

The condensate is treated with a nonlinear Schr\"odinger equation (NLSE)
within the Popov approximation.
The thermal atoms are described using the Popov equations which are eigenequations.
Because of computational complexity, we use the Popov equations only for excitations 
below a certain cutoff energy $E_{\mathrm{cut}}$.
The excitation energies $\varepsilon > E_{\mathrm{cut}}$ are
taken into account
within the semiclassical approximation~\cite{giorgini,dalfovo,reidl}.
The density of the condensate is $n_0({\bm r}) \equiv |\phi({\bm r})|^2$, where $\phi({\bm r})$ is
the condensate wavefunction.
The normal particle density coming from the excitations below (above) $E_{\mathrm{cut}}$
is $n_1$ ($n_2$).
Therefore, the particle number density is 
$n({\bm r}) = n_0({\bm r}) + n_1({\bm r}) + n_2({\bm r})$.
The condensate is described with the NLSE 
\begin{equation}
    \{
        - C \nabla^2 + V - \mu
        + g(n_0 + 2n_1 + 2n_2)
    - {\bm{\omega}}_\mathrm{rot}\cdot{\bm r}\times {\bm p}
\} \phi  =  0
\label{eq:gp}
\end{equation}
where
$C = \hbar^2/(2m)$ and $g = 4\pi\hbar^2a/m$.
The mass of a Na atom $m = 38.17 \times 10^{-27} \text{ kg}$,
and the scattering length $a = 2.75 \text{ nm}$ for Na atoms 
are employed.
We use the cutoff energy $E_{\mathrm{cut}} = 10 \, \hbar \omega_\mathrm{tr}$ \cite{cutoff10}.
Angular velocity of rotation is $\omega_\mathrm{rot}$ and the rotation axis is
parallel with the $z$ axis ($\bm{\omega}_\mathrm{rot} = {\bm e}_z \omega_\mathrm{rot}$).
Excitations below the cutoff $E_{\mathrm{cut}}$ are eigenstates of the Popov equations
%
%
%------------------------------
%
\begin{subequations} \label{eq:popov} \begin{eqnarray}
\lefteqn{
    \{
        -C \nabla^2 + V - \mu + 2g(n_0 + n_1 + n_2)
}\nn\\*
&&
    - {\bm{\omega}}_\mathrm{rot} \cdot {\bm r} \times {\bm p} \}u_q
    - g\phi^2 v_q = \varepsilon_q u_q ,
\label{eq:popov1}\\
\lefteqn{
    -\{
        -C \nabla^2 + V - \mu + 2g (n_0 + n_1 + n_2)
}\nn\\*
&&
    + {\bm{\omega}}_\mathrm{rot} \cdot {\bm r} \times {\bm p} \}v_q
   + g\phi^{\ast 2} u_q = \varepsilon_q v_q.
    \label{eq:popov2}
\end{eqnarray} \end{subequations}
These reduce to the Bogoliubov equations if we neglect $n_1$ and $n_2$.
The wavefunctions $u_q$ and $v_q$ obey the normalization condition
\begin{eqnarray}
\int \left( |u_q|^2 - |v_q|^2 \right) d{\bm r} = 1.
\end{eqnarray}
The density $n_1$ is a weighted sum of the wavefunctions $u$ and $v$:
\begin{eqnarray}
    n_1({\bm r}) &=& 
        \left\{
            \sum_{q\, (\varepsilon_q < E_{\mathrm{cut}})}
            \left( |u_q|^2 + |v_q|^2 \right) f(\varepsilon_q) + |v_q|^2
        \right\},
    \label{eq:rho_1}
    \\
    f(\varepsilon) &=& \frac{1}{e^{\varepsilon / (k_{B}T)} - 1}.
\end{eqnarray}
The higher-energy range $\varepsilon > E_\mathrm{cut}$ is described within
the semiclassical approximation~\cite{giorgini,dalfovo,reidl}
which neglects the derivatives of the amplitudes of the wavefunctions
$u$ and $v$ and the second derivatives of their phases.
We also neglect the phase of the condensate wavefunction $\phi$ here.
Then the Popov equations (\ref{eq:popov}) reduce into 
algebraic form (Ref.~\cite{reidl}, Eqs. (5)).
The expression for $n_2$, in analogy with Eq.~(\ref{eq:rho_1}), is:
\begin{equation}
    n_2({\bm r}) = \int \frac{d{\bm p}}{h^3}
    \left\{
        \frac{\varepsilon_\mathrm{HF}}{\tilde{\varepsilon}}
        \left(
	    f(\varepsilon) + \frac{1}{2}
        \right)
        - \frac{1}{2}
    \right\}
%\nn \\ &&
    \Theta\left( \varepsilon - E_{\mathrm{cut}} \right)
       \label{eq:ldbog}
\end{equation}
where the Hartree-Fock (HF) energy
\begin{equation}
   \varepsilon_{\mathrm{HF}}({\bm r}, {\bm p}) = 
       \frac{{\bm p}^2}{2m} + V - \mu + 2g(n_0 + n_1 + n_2)
       \label{eq:ehf}
\end{equation}
and energies
\begin{eqnarray}
   \tilde{\varepsilon}({\bm r}, {\bm p})  &=& 
    \sqrt{ \varepsilon_\mathrm{HF}^2({\bm r}, {\bm p}) - g^2 n_0 },
       \label{eq:etilde}
\\
   \varepsilon({\bm r}, {\bm p})  &=& \tilde{\varepsilon}({\bm r}, {\bm p})
        - {\bm{\omega}}_\mathrm{rot} \cdot {\bm r} \times {\bm p}
       \label{eq:e}
\end{eqnarray}
are functions of ${\bm r}$ and ${\bm p}$.
The noncondensate densities $n_1$ and $n_2$
are determined from Eqs.~(\ref{eq:rho_1}) and (\ref{eq:ldbog}).
They are treated as mean field potentials throughout the above equations.
Thus the numerical procedure needs to be 
selfconsistent such that it is
repeated until
the solution reaches convergence in which all the equations are simultaneously satisfied.

\begin{figure}[tb]
\begin{center}
\includegraphics[width=8cm]{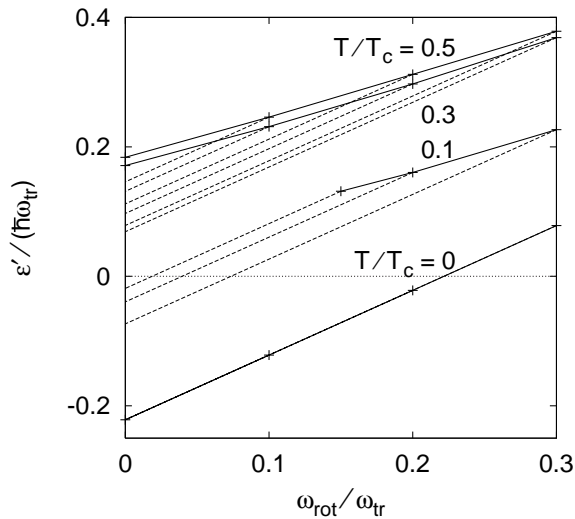}
\end{center}
\caption{\label{fig:axi}
Dependence of $\varepsilon^\prime$ (lowest $\varepsilon$) on $\omega_{\mathrm{rot}}$
in an axisymmetric system ($\Delta r = 0$).
The solid lines indicate the $\varepsilon^\prime$ at $T = 0$ (Bogoliubov approximation) and
$T/T_c = 0.1, 0.3, \text{ and } 0.5$ (Popov approximation).
The dotted lines represent Eq.~(\ref{eq:e_lab}).
The angular momentum $q_\theta = -1$.
The dotted line and the solid line overlap at $T=0$.
The two lines are separated for $T>0$, which means that ``equality"
between the rotation velocity $\omega_\mathrm{rot}$ and 
the excitation energy in Eq.~(\ref{eq:e_lab}) is lost at finite temperatures.
}
\end{figure}

The angular momenta of the condensate $\phi$ and of the wavefunctions $u, v$ are
\begin{eqnarray}
    \mathcal{A}(\phi)
    &=&
    {\bm e_z}\cdot
    \frac{
        \int \phi^{\ast}({\bm r} \times {\bm p})\phi \, d{\bm r}
    }{
        \hbar \int n_0({\bm r}) d{\bm r}
    }, 
\label{eq:L} \\  
    \mathcal{A}(u_q) &=&
    {\bm e_z}\cdot
    \frac{
        \int u_q^{\ast}({\bm r} \times {\bm p})u_q \, d{\bm r}
    }{
        \hbar U_q
    },
\label{eq:QThetaU} \\  
    \mathcal{A}(v_q) &=&
    {\bm e_z}\cdot
    \frac{
        \int v_q^{\ast}({\bm r} \times {\bm p})v_q \, d{\bm r}
    }{
        \hbar V_q
    } 
\label{eq:QThetaV}
\end{eqnarray}
where $U_q \equiv \int |u_q|^2 d{\bm r}$ and $V_q \equiv \int |v_q|^2 d{\bm r}$. 
Within the Bogoliubov theory ($T = 0$), the excitation energies
depend linearly on the angular velocity $\omega_{\mathrm{rot}}$ as follows:
\begin{eqnarray}
    \varepsilon &=& \varepsilon_{\mathrm{lab}} - \hbar \omega_{\mathrm{rot}} q_{\theta},
\label{eq:e_lab} \\
    q_{\theta} &\equiv& 
    \mathrm{Re}\left[
        \{ \mathcal{A}(u_q) - \mathcal{A}(\phi) \} U_q +
        \{ \mathcal{A}(v_q) + \mathcal{A}(\phi) \} V_q
    \right]
\nn\\*
\lefteqn{
      / (U_q + V_q). 
}
\label{eq:q_theta} 
\end{eqnarray}
The value $q_{\theta}$ is useful for characterizing~\cite{baksmaty0307368,mizushima}
an excitation also for finite-temperature systems.
But as for the $\omega$ dependence of $\varepsilon$,
there are deviations from Eq.~(\ref{eq:e_lab}) due to
changes of density in the normal component
$n_1({\bm r}) + n_2({\bm r})$ for finite-temperature systems.
Figure \ref{fig:axi} indicates these deviations.
Changes in the excitation energies modify the normal-component density
which in turn
affects the whole system, including the $\varepsilon$'s themselves.
Therefore, the spectra for each value of $\omega_{\mathrm{rot}}$
needs to be calculated individually.

%----------------------------------------
%----------------------------------------

%----------------------------------------
\begin{figure}[tb]
\begin{center}
\includegraphics[width=8cm,clip]{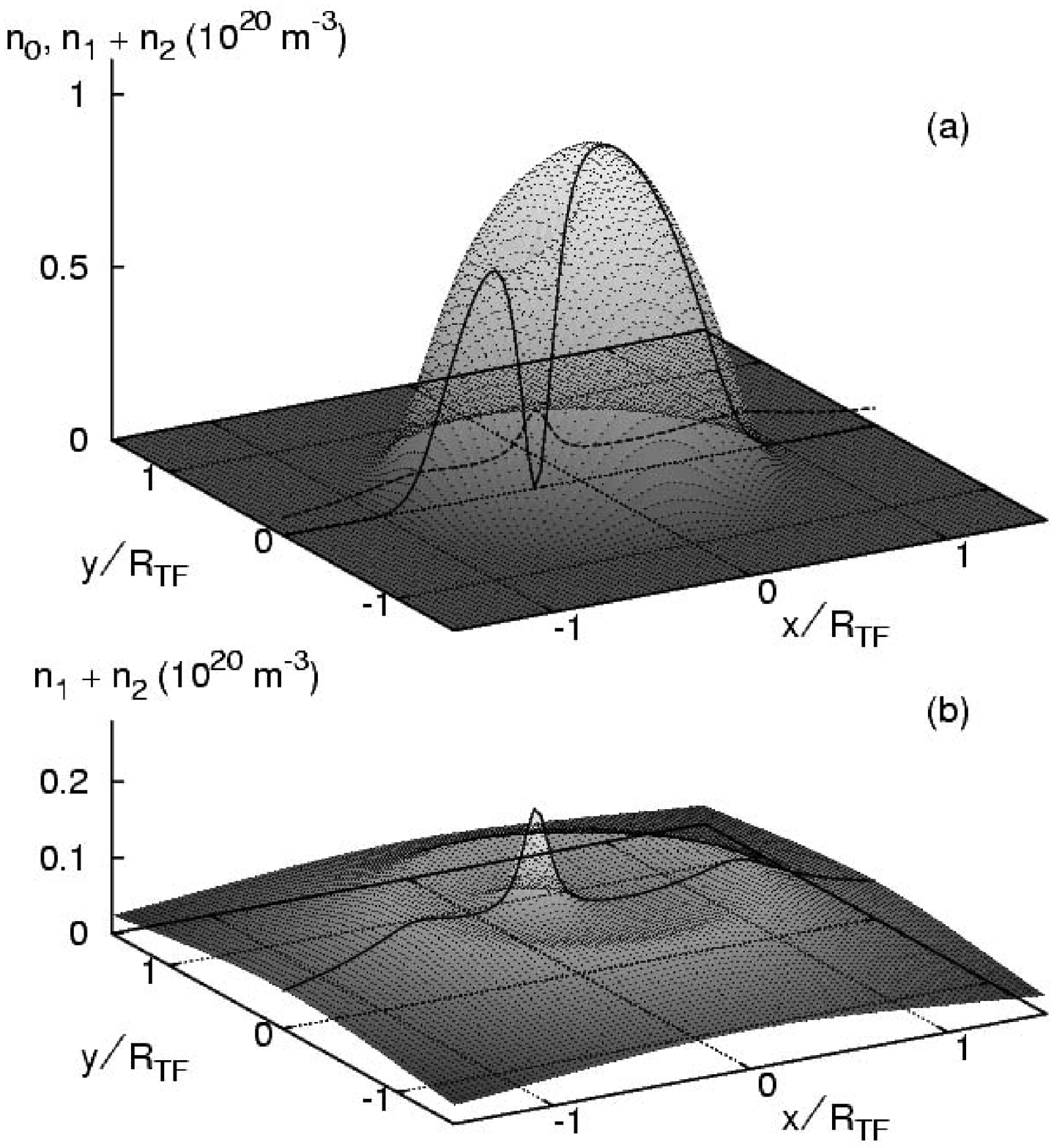}\\
\includegraphics[width=7cm]{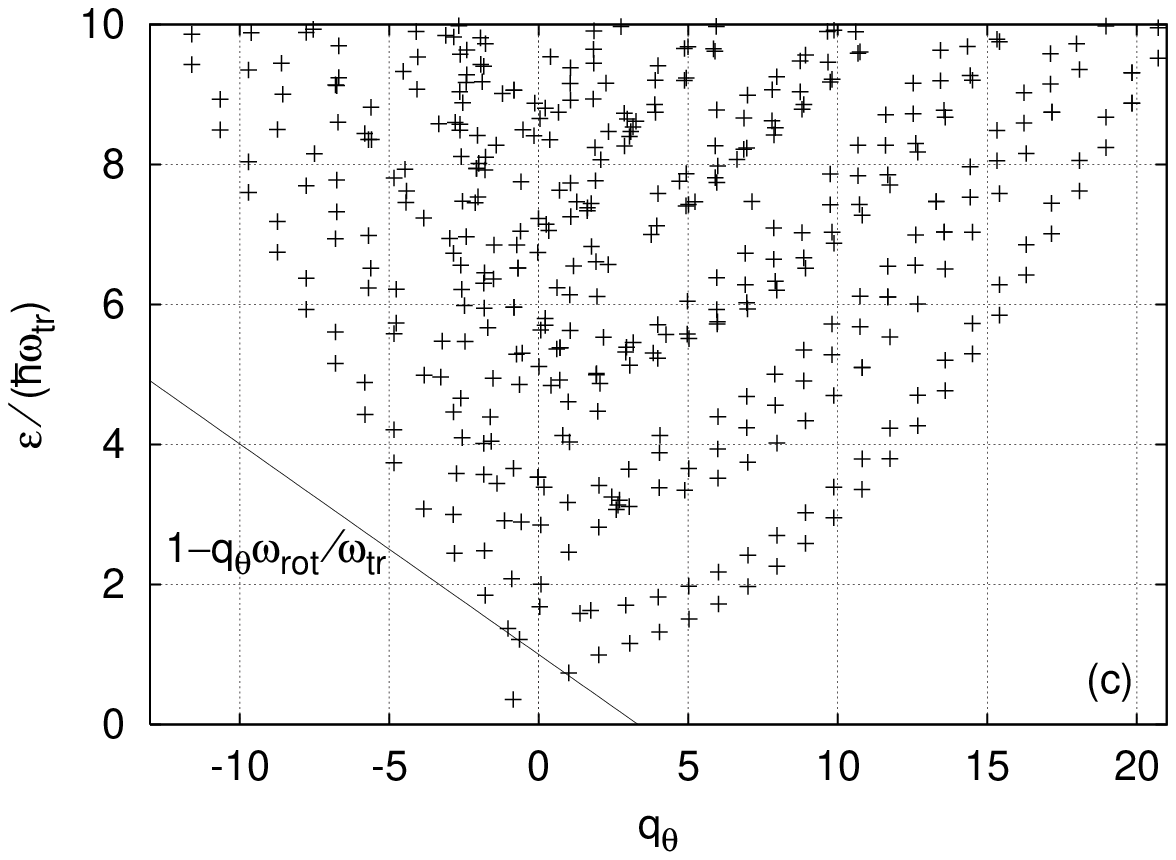}
\end{center}
\caption{\label{fig:dns}
Density profiles of a system at the temperature $T = 0.5\, T_c$, and
angular momentum of the condensate $\mathcal{A}(\phi) =  0.863$ in 
rotating coordinates
with $\omega_{\mathrm{rot}} = 0.300\, \omega_{\mathrm{tr}}$.
(a) Particle density $n_0(x,y)$ of the condensate,
(b) Particle density $n_1(x,y) + n_2(x,y)$  of the noncondensate,
(c) Excitation spectra $\varepsilon_q$ vs. $q_{\theta}$. The solid line shows
$1 - \omega_{\mathrm{rot}}q_{\theta}/\omega_{\mathrm{tr}}$.
Its slope indicates the rotation velocity  $\omega_{\mathrm{rot}}$.
The line touches the dipole modes at $q_\theta = \pm 1$.
}
\end{figure}
%----------------------------------------

%======================================================
\section{Off-centered vortex}

The sign of the excitation energy of the core-localized excitation and
the direction of the precessional motion are related~\cite{linn} at zero temperature.
The predicted precession frequency fits well with results of the experiments~\cite{feder}.
A finite-temperature extension of the Bogoliubov equation,
the Popov approximation, shows
that the sign of the core-localized excitation becomes positive~\cite{tomoya,tapsa}.
If the direction of the precessing motion and the
sign of the excitation energy correspond to each other,
the precessing motion of the vortex must also be inverted 
as the lowest excitation energy raises from
negative to positive values.

We extend the 2D treatment~\cite{jukka} of the Bogoliubov equations to
finite-temperature Popov equations.
It makes possible to treat the slightly off-centered vortices 
directly within the Popov approximation.

Assume a BEC system has a vortex line perpendicular to the $z$ axis.
Particles are confined with a harmonic trap along the $x$ and $y$ axes
\begin{equation}
V(x,y) = \frac{m\omega_\mathrm{tr}^2}{2}(x^2 + y^2)
\end{equation}
with $\omega_\mathrm{tr} = 2\pi \times 200 \, \mathrm{Hz}$.
The system has finite $z$ thickness and it is uniform along the $z$ axis.
Periodic boundary conditions along $z$ are used.
Hence this system is not a two-dimensional BEC~\cite{pricoupenko,gies},
but rather a three-dimensional BEC having restricted geometry.
We treat a system having $10^5$ atoms within a $z$ thickness of $10 \, \mu \mathrm{m}$.
The Thomas-Fermi (TF) radius $R_{\mathrm{TF}} = 6.793\,\mu \mathrm{m}$ is used 
as the scale of length.

Equations (\ref{eq:gp}) - (\ref{eq:e})  are repeatedly solved
until convergence into a self-consistent solution.
While the vortexfree $(\mathcal{A}(\phi) = 0)$ and centered-vortex $(\mathcal{A}(\phi) = 1)$
configurations are most likely, there also exists a solution with an off-centered vortex
$(0< \mathcal{A}(\phi) < 1)$ in a narrow window of $\omega$.
Figures \ref{fig:dns}(a-c)
represent such a typical system. Here the angular momentum is $\mathcal{A}(\phi) = 0.863$.
The particle number of the condensate is $48\%$ of the total particle number.
The noncondensate density has a characteristic peak at the core of the vortex,
like those in the axisymmetric studies~\cite{tomoya,tapsa}.

\begin{figure}
\begin{center}
\includegraphics[width=8.5cm]{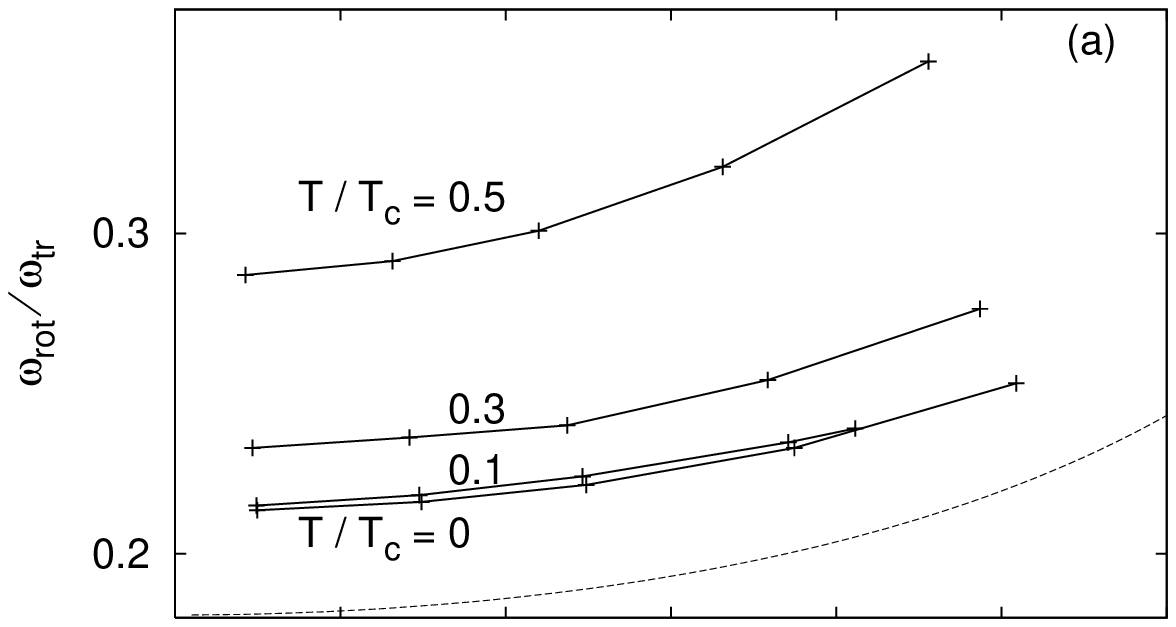}\\

\vspace{-0.4cm}
\includegraphics[width=8.5cm]{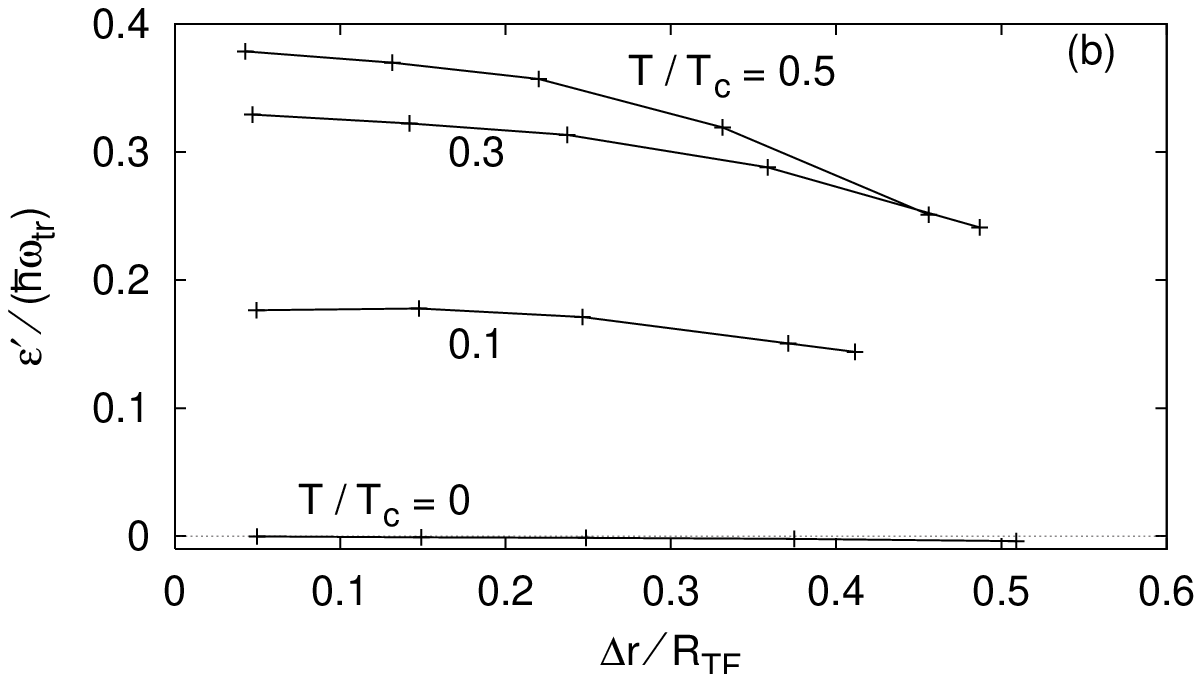}\\
\end{center}
\caption{\label{fig:rtf}
Horizontal axis indicates the displacement $\Delta r$ of the vortex core
from the center of the harmonic trap.
(a) The vertical axis is the angular velocity $\omega_{\mathrm{rot}}$ of the rotating coordinates
for temperatures $T = 0.0, 0.1, 0.3, \text{ and } 0.5\,  T_c$.
The difference in the frequencies is less than $2\%$ between a point at $T=0$ and
a corresponding one at $T=0.1\, T_c$ within $\Delta r < 0.4$.
The dotted line is the result of the TF approximation at $T=0$~\cite{lundh:ao}.
Our results for $T=0.1\, T_c$ are $16 - 19\%$ above those of the TF approximation.
(b) The lowest excitation energy.
All of the energies in the finite-temperature ($T = 0.1, 0.3 \text{ and } 0.5$) systems
are positive.
The energy of the lowest core localized excitation only
turns negative for $T=0$ when the normal densities $n_1$ and $n_2$ are neglected.
}
\end{figure}

The displacement of the vortex core $\Delta r$ is unrestricted
in the numerical processes, unlike for the axisymmetric situations $(\Delta r = 0)$.
Therefore, the displacement $\Delta r$ depends on
temperature $T$ and the rotation frequency $\omega_{\mathrm{rot}}$ 
as presented in Fig.~\ref{fig:rtf}(a).

Figure \ref{fig:rtf}(a) plots rotation frequencies at which the system is static.
Let us denote the displacement and the rotation frequency at the static point 
as $\Delta r^{\prime}$ and $\omega_\mathrm{rot}^{\prime}$.
When $\Delta r < \Delta r^{\prime}$, $\omega_\mathrm{rot} = \omega_\mathrm{rot}^{\prime}$, and $T=0$,
the vortex has an instability and it tends to move inward.
When $\Delta r > \Delta r^\prime$, the direction is outward.
As $\omega_\mathrm{rot}$ increases in Fig.~\ref{fig:rtf}(a),
the system has a wider range of displacements $\Delta r$ having an inward instability.
This instability brings the vortex to the center of system and makes the vortex state more sustainable.
%%%%%%%%%%%%%%%%%%%%%%% Mar 08 2004
At the finite temperature $T=0.1 T_c$,
the range of $\Delta r$ having an inward instability is almost the same.
But the positive value of the lowest excitation energy $\varepsilon^\prime$
in Fig.~\ref{fig:rtf}(b)
requires that the displacement $\Delta r$ is stable and the vortex tends to
remain at $\Delta r = \Delta r^{\prime}$.

The lowest excitation energies $\varepsilon^\prime$ in Fig.~\ref{fig:rtf}(b)
differ significantly between the $T=0$ system and those of the $T=0.1\, T_c$ system.
It is small and negative ($0 > \varepsilon^\prime > - 0.004 \hbar \omega_\mathrm{tr}$)
for the $T=0$ system, while it is positive in the $T=0.1\, T_c$ system.
The corresponding rotation frequencies $\omega_{\mathrm{rot}}$
in Fig.~\ref{fig:rtf}(a) differ only a little.
This indicates that the direction of the precessional motion
which is represented by the sign of $\omega_{\mathrm{rot}}$
and the sign of the lowest core excitation energy are not related in the $T>0$ systems.
This point is discussed further in the next section.

\begin{figure}
\begin{center}
\includegraphics[width=8.5cm]{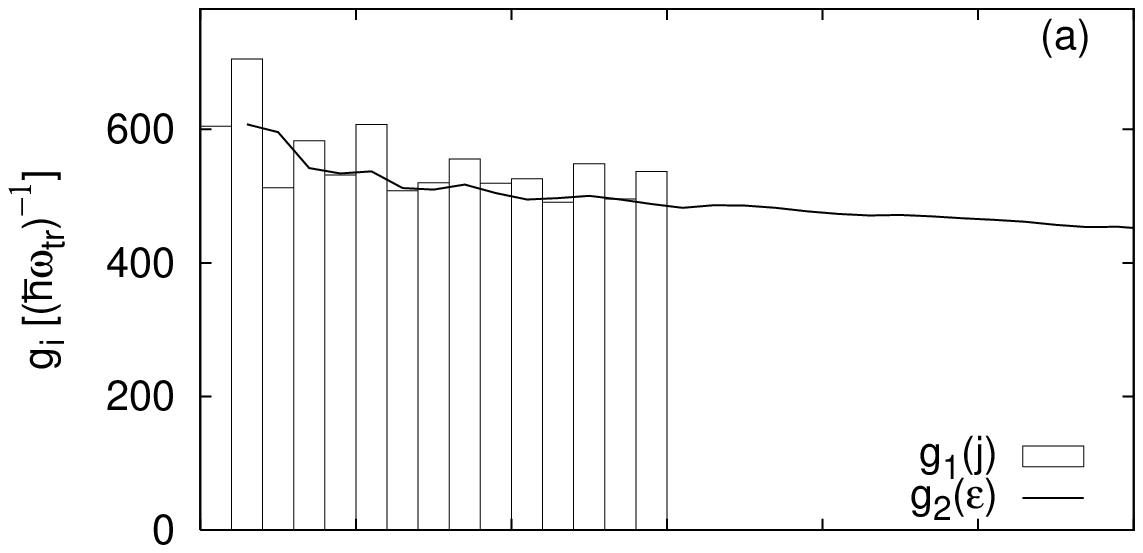}

\vspace{-0.4cm}
\includegraphics[width=8.5cm]{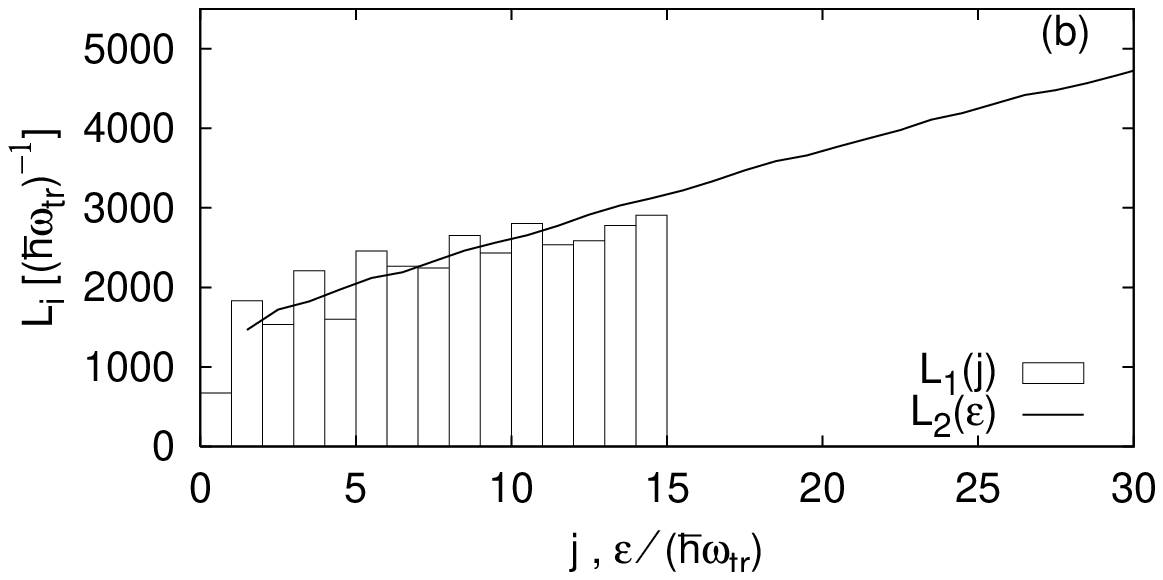}
\end{center}
\caption{\label{fig:spect} 
(a) Spectral distributions $g_1(\varepsilon)$
obtained within the Popov approximation and
$g_2(\varepsilon)$ obtained within the semiclassical approximation.
(b) Angular-momentum distributions $L_1$ and $L_2$.
These two approximations yield consistent
particle numbers
and angular momenta for a wide range of energies.
The temperature $T/T_c = 0.5$ and the angular momentum $\mathcal{A}(\phi) = 0.863$.
}
\end{figure}

Figures \ref{fig:spect}(a-b) compare the spectral densities
\begin{eqnarray}
g_1(j) &=&
    \int d{\bm r}\!\!\!\!\!
    \sum_{q, j \le  \frac{\varepsilon_q}{\hbar\omega_\mathrm{tr}}< (j+1)}
    \!\!\!\!\!
    \left( |u_q|^2 + |v_q|^2 \right) f(\varepsilon_q) + |v_q|^2, 
\nn \\* &&
    \quad \label{eq:g1}
\\
g_2(\varepsilon^{\prime\prime}) \! &=&  \!
    \int \!\! \frac{d{\bm p} d{\bm r}}{h^3}
    \left\{
        \frac{\varepsilon_\mathrm{HF}}{ \tilde{\varepsilon} }
        \left(
	    f(\varepsilon) + \frac{1}{2}
        \right)
        - \frac{1}{2}
    \right\}
    \delta(\varepsilon^{\prime\prime} - \varepsilon) 
\end{eqnarray}
and the angular momenta of the noncondensate
\begin{eqnarray}
    L_1(j) &=&
    {\bm e_z}\cdot
    \frac{1}{\hbar}
    \sum_{q, j \le  \frac{\varepsilon_q}{\hbar\omega_\mathrm{tr}}< (j+1)}
    \int d{\bm r} [
        \{
       	    u^{\ast}_q({\bm r} \times {\bm p})u_q
\nn\\&&
	    +
            v_q({\bm r} \times {\bm p})v_q^{\ast} 
        \}
        f(\varepsilon_q)
	+ 
            v_q({\bm r} \times {\bm p})v_q^{\ast} ],
\\
    L_2(\varepsilon^{\prime\prime}) &=&
    {\bm e_z}\cdot
    \frac{1}{\hbar}\int \frac{d{\bm p} d{\bm r}}{h^3}
    \left( {\bm r}\times {\bm p} \right)
    \{
        f(\varepsilon({\bm r},{\bm p})) 
\nn\\&&
    - \frac{1}{2}
    \left(
       \frac{\varepsilon_{\mathrm{HF}}({\bm r},{\bm p})}{\tilde{\varepsilon}({\bm r},{\bm p})}
    - 1
    \right)
    \}
    \times\delta(\varepsilon^{\prime\prime} - \varepsilon)
       \label{eq:L2}
\end{eqnarray}
to verify the mutual consistency of the Popov approximation
and that of the semiclassical approximation.
Above,
$g_1$ and $L_1$ are obtained within the Popov approximation, while
$g_2$ and $L_2$ are obtained within the semiclassical approximation.
Results of these two approximations are consistent with each other
for $\varepsilon > 5 \, \hbar \omega_\mathrm{tr}$.

The core-localized mode is mainly affected by the particle densities inside the core,
while $g_1$ and $g_2$ account for the densities over the whole area of the system. 
Since our main interest is the core-localized mode,
we employ calculations with reduced accuracy
for $n_2$ and $g_2$ for larger $x, y$ (around $|x| > 1.2 R_{\mathrm{TF}}$ or $|y|> 1.2 R_{\mathrm{TF}}$).
This affects the plot for $g_i$ in Fig.~\ref{fig:spect}(a),
but it has little effect on $\varepsilon^{\prime}$
as shown in the dependence \cite{cutoff10} of $\varepsilon^{\prime}$ on the cutoff energy,
$E_\mathrm{cut}$.

%----------------------------------------
%----------------------------------------
%----------------------------------------
%----------------------------------------

%======================================================
\section{The sign and direction of vortex precession}

The angular velocity $\omega_\mathrm{prec}$ of the precessional motion 
of a vortex
is related with that of the rotating frame $\omega_\mathrm{rot}$  
through
\begin{equation}
    \omega_\mathrm{prec} + \omega_\mathrm{rot} = \text{const.}  \label{eq:prec:rot}
\end{equation}
within the GP equations.
The core-localized excitation has the lowest excitation energy $\varepsilon^{\prime}$
within the range of $\omega_\mathrm{rot}$ we treat.
Using variational Lagrangean analysis \cite{linn}, it can be shown that
\begin{equation}
    \varepsilon^{\prime} = 0 \text{ at } \omega_\mathrm{prec} = 0 \label{eq:ezero}
\end{equation}
for a small displacement $\Delta r$.
This relation remains valid for $0 < \Delta r < 0.5 R_{\mathrm{TF}}$ within
the accuracy $0 > \varepsilon^{\prime} > -0.004 \hbar\omega_{\mathrm{tr}}$
in the present system, see Fig.~\ref{fig:rtf}(b).

Equations (\ref{eq:e_lab}), (\ref{eq:prec:rot}), and (\ref{eq:ezero})
lead to the relation
\begin{equation}
    \varepsilon^{\prime} = \hbar q_{\theta} \omega_\mathrm{prec} \label{eq:e:and:prec}.
\end{equation}
Therefore, the direction of the precessional motion
and the sign of the core-localized excitation correspond to each other at $T=0$.

If the coordinate transformation in Eq.~(\ref{eq:prec:rot}) were not valid,
the relation Eq.~(\ref{eq:e:and:prec})  between 
the excitation energy $\varepsilon^{\prime}$ and 
the angular velocity $\omega_\mathrm{prec}$ of the precessional motion
would not hold.
The next section describes how this occurs at finite temperatures.

An easier way to disprove Eq.~(\ref{eq:e:and:prec}) is as follows.
Figure \ref{fig:rtf} (b) displays $\varepsilon^{\prime}$
for $\omega_\mathrm{prec} = 0$.
It shows that Eq. (\ref{eq:ezero}) is no longer satisfied for $T>0$.
Therefore, it becomes impossible to satisfy Eq.~(\ref{eq:e:and:prec}).
The sign of the lowest excitation and
the direction of the precessional motion
are no longer related within the finite-temperature (Popov) approximation.

%====================================================
\section{Precessing and Rotating Frames}\label{sec:adiab}

We have presented density profiles of an off-centered vortex within the rotating frame.
This frame mimics a system of atoms in a rotating trap
although the deformation of the trapping potential is not included explicitly.
The vortices are precessing if we observe them from the nonrotating frame.
But there is a restriction in the velocity of the precession.

\begin{figure}
\begin{center}
\includegraphics[width=8.5cm]{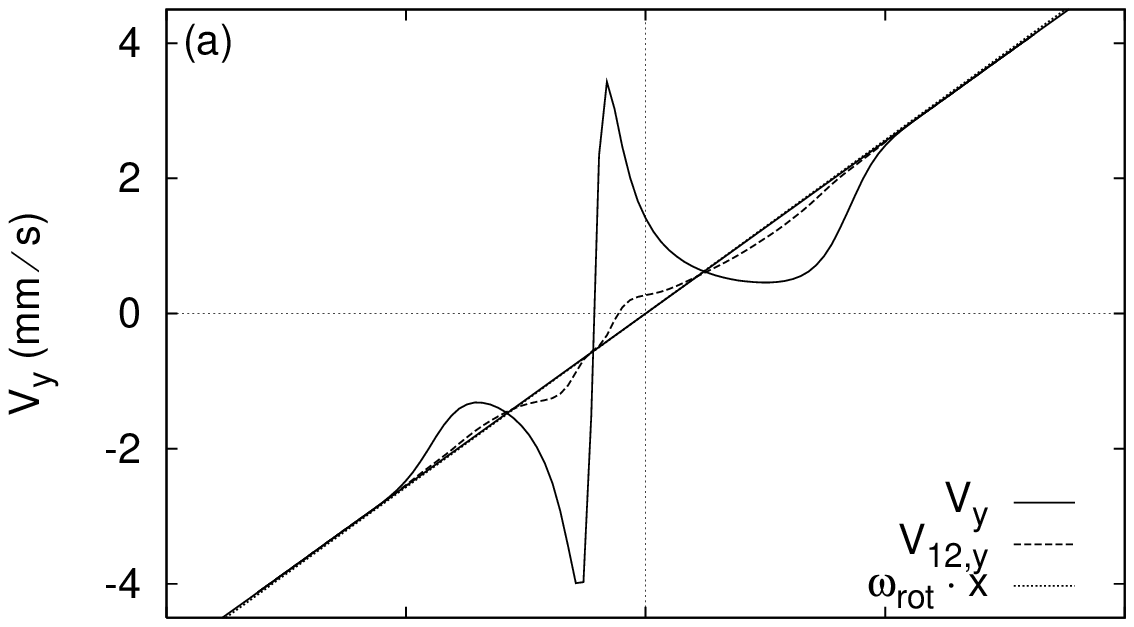}
\vspace{-0.4cm}

\includegraphics[width=8.5cm]{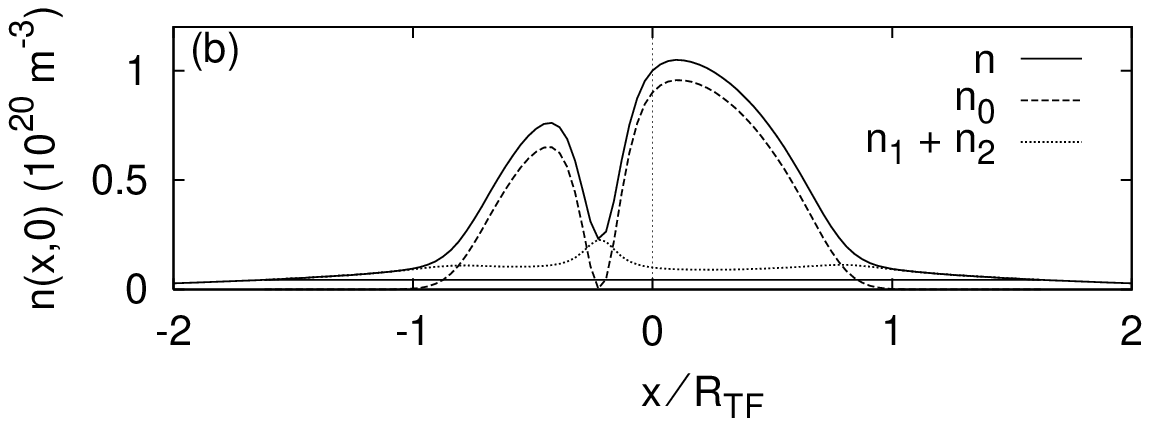}\\
\end{center}
\caption{\label{fig:mvn} 
(a) Velocities of all the atoms (solid line) and the normal component (dashed line)
along the $x$ axis $(y = 0)$.
The velocity follows that of the rotating frame (dotted line) outside
the condensate ($|x|>R_{\mathrm{TF}}$).
Temperature $T = 0.5  T_c$ and angular momentum $\mathcal{A}(\phi) = 0.863$.
(b) Density profiles along the $x$ axis for comparison with (a).
The plots of $n_0$ (dashed line) and $n_1 + n_2$ (dotted line) are
equivalent to those on the $x$ axes in Figs.~\ref{fig:dns}(a-b).
}
\end{figure}

Let us compare the velocities of the noncondensate and the angular
velocity of the frame $\omega_\mathrm{rot}$ at $\omega_\mathrm{prec}=0$.
Figure \ref{fig:mvn} (a) displays the local velocities on the $x$ axis for the system.
The velocities  
\begin{eqnarray}
  {\bm V}_{12}({\bm r}) &=&
   \frac{
      \mathrm{Re}[{\bm P}_1({\bm r}) + {\bm P}_2({\bm r})]
   }{
      m \left\{ n_1({\bm r}) + n_2({\bm r}) \right\}
   },
\\
   {\bm V}({\bm r}) &=& \frac{
      \mathrm{Re}[{\bm P}_0({\bm r}) + {\bm P}_1({\bm r}) + {\bm P}_2({\bm r})]
   }{
      m n({\bm r})
   }
\end{eqnarray}
which are particle-current densities divided by the particle densities;
they are defined using the momentum densities
\begin{eqnarray}
    {\bm P}_0({\bm r})
    &=& \phi^{\ast}({\bm r}) {\bm p} \phi({\bm r}),
\\
    {\bm P}_1({\bm r})
    &=&
    \sum_{q\, (\varepsilon_q < E_{\mathrm{cut}})}
        \left\{
       	    u^{\ast}_q({\bm r}) {\bm p} u_q({\bm r}) + v_q({\bm r}) {\bm p} v_q^{\ast}({\bm r}) 
        \right\}
        f(\varepsilon_q)
\nn\\&&
	+ 
            v_q({\bm r}) {\bm p} v_q^{\ast}({\bm r}),
\\
    {\bm P}_2({\bm r})
    &=&
    \int \frac{d{\bm p}}{h^3} {\bm p}
    \left\{
        f(\varepsilon({\bm r},{\bm p})) 
    - \frac{1}{2}
    \left(
       \frac{\varepsilon_{\mathrm{HF}}({\bm r},{\bm p})}{\tilde{\varepsilon}({\bm r},{\bm p})}
    - 1
    \right)
    \right\}\times
\nn\\&&
    \Theta\left[ \varepsilon({\bm r},{\bm p}) - E_{\mathrm{cut}} \right].
\end{eqnarray}
The velocity closely follows that of the rotating frame outside
of the condensate ($|x|>R_{\mathrm{TF}}$) in Fig.~\ref{fig:mvn}(a).
It is straightforward to show that
${\bm V}({\bm r})$ reduces to
\begin{equation}
    {\bm V}({\bm r}) = \omega_{\mathrm{rot}}(-y, x, 0)
\end{equation}
for large $|{\bm r}|$ where the density of the condensate and
normal component below the cutoff are negligible ($n_0 + n_1 \ll n$).
This indicates adiabaticity between the rotating trap and the normal gas.
Within the Popov approximation, the angular velocity 
of the precessional motion of a vortex
is restricted to that of a normal gas and a confining trap.

We have considered static systems with $\omega_\mathrm{prec} = 0$
in a rotating frame with the angular velocity $\omega_\mathrm{rot}$.
These two $\omega$'s may be transformed between each other
using the simple relationship between 
the stationary and rotating frames of reference
in Eq.~(\ref{eq:prec:rot}), valid at zero temperature.
But taking into account the normal component,
this relation is only valid when the normal component is rotating
at the angular velocity $\omega_\mathrm{rot}$.
Varying $\omega_\mathrm{rot}$ will change the density profiles
through Eqs.~(\ref{eq:e}) and (\ref{eq:e_lab}).
Then the coordinate transformation Eq.~(\ref{eq:prec:rot}) does not hold.
Adiabaticity required within the Popov framework
restricts the recognizing of $\omega_\mathrm{rot}$ as 
the angular velocity of precession $\omega_\mathrm{prec}$.
This is another reason why Eq.~(\ref{eq:e:and:prec}) does not hold
at finite temperatures.

%----------------------------------------
%----------------------------------------

%======================================================
\section{Discussion}

It is confirmed that the sign of the lowest excitation energy $\varepsilon^\prime$
is, in general, unrelated with the direction of the precessional motion of a vortex
within the Popov approximation.
Within the Bogoliubov approximation,
the excitation energy $\varepsilon^\prime$ of core-localized mode and
the precession frequency $\omega_\mathrm{prec}$
are proportional to each other as shown in Eq.~(\ref{eq:e:and:prec}).
The derivation of Eq.~(\ref{eq:e:and:prec}) shows that the angular velocity
(and the direction) $\omega_\mathrm{prec}$
of the precessional motion arises from
the coordinate transformation Eq.~(\ref{eq:prec:rot}),
which does not have any explicit relation with the core-localized excitation. 
Therefore, the core-localized excitation is responsible for the inward/outward motions
of vortex and not explicitly related to the precessional motion.

We think that this nature of the core-localized excitation
does not change even at a finite $T$.
However, a system described within the Popov approximation
cannot undergo the coordinate transformation.
Within the Popov approximation, 
the velocity $\omega_\mathrm{rot}$ in Fig.~\ref{fig:rtf}(a) may
be recognized as the angular velocity of
the precessional motion only when the condition of adiabaticity is obeyed.
Therefore, the relation between the sign of
the lowest excitation energy $\varepsilon^\prime$
and the sign of $\omega_\mathrm{prec}$, \textit{i.e.},
the direction of the precessional motion of a vortex is broken 
within the Popov approximation.

We obtained the precession frequencies of a vortex at finite temperatures.
The vortex-precession frequency manifests [Fig.~\ref{fig:rtf}(a)] the same tendency
as a function of displacement $\Delta r$ as the
zero-temperature TF studies~\cite{fetter:review,lundh:ao}.
The precession frequency
increases as the displacement $\Delta r$ and the temperature increase.
The frequencies differ only little between those for $T=0$ and $T=0.1\, T_c$,
despite the large changes in the excitation spectra.
Although a pinning effect \cite{tomoya,tapsa,pricoupenko} caused by the
normal component makes a significant difference
in the excitation energy $\varepsilon^\prime$,
it has little effect on the precession frequencies between
the systems at zero and finite temperatures.

%======================================================
\section*{Acknowledgements}

The authors are grateful to G. Baym, E. Lundh, M. M\"{o}tt\"{o}nen, and S. M. M. Virtanen for stimulating discussions.
One of the authors (T.I.) is supported by the bilateral exchange program between the
Academy of Finland and JSPS, the Japan Society for the Promotion of Science.

%======================================================


\begin{thebibliography}{99}
\bibitem{cutoff10}
The choice for the cutoff has very little effect on the energy $\varepsilon^\prime$
of the core-localized mode.
At $T / T_c = 0.5$ and $A(\phi)=0.863$,
the energies $\varepsilon^\prime$ for various cutoff values are
$(E_{\mathrm{cut}}/(\hbar \omega_\mathrm{tr}), \varepsilon^\prime/(\hbar \omega_\mathrm{tr})) = (5, 0.345), (10, 0.354),$ and $(20, 0.357).$
\end{thebibliography}
\end{document}